# GPA: Grover Policy Agent for Generating Optimal Quantum Sensor Circuits

Ahmad Alomari and Sathish A. P. Kumar
Department of Computer Science, Cleveland State University, Cleveland, OH USA, 44115
a.alomari@vikes.csuohio.edu; s.kumar13@csuohio.edu

*Abstract*—This study proposes a GPA for designing optimal Quantum Sensor Circuits (QSCs) to address complex quantum physics problems. The GPA consists of two parts: the Quantum Policy Evaluation (QPE) and the Quantum Policy Improvement (QPI). The QPE performs phase estimation to generate the search space, while the QPI utilizes Grover search and amplitude amplification techniques to efficiently identify an optimal policy that generates optimal QSCs. The GPA generates QSCs by selecting sequences of gates that maximize the Quantum Fisher Information (QFI) while minimizing the number of gates. The QSCs generated by the GPA are capable of producing entangled quantum states, specifically the squeezed states. High QFI indicates increased sensitivity to parameter changes, making the circuit useful for quantum state estimation and control tasks. Evaluation of the GPA on a QSC that consists of two qubits and a sequence of $R_x$, $R_y$, and $S$ gates demonstrates its efficiency in generating optimal QSCs with a QFI of 1. Compared to existing quantum agents, the GPA achieves higher QFI with fewer gates, demonstrating a more efficient and scalable approach to the design of QSCs. This work illustrates the potential computational power of quantum agents for solving quantum physics problems.

*Impact Statement*—The GPA introduces an AI-driven approach to quantum circuit optimization, advancing quantum sensing and metrology. By leveraging QPE and QPI, the GPA efficiently generates QSCs with high QFI while minimizing gate complexity and enhancing quantum state estimation and precision measurement. Technologically, the GPA advances AI-driven quantum optimization, making quantum sensing and metrology more scalable and efficient. Economically, it improves quantum computing efficiency, lowers operational costs, and accelerates commercialization. Socially, optimized QSCs benefit biomedical imaging, secure quantum communication, and geophysics. Legally, the findings contribute to standardizing quantum metrology for scientific and industrial applications. By demonstrating a pure quantum agent that does not use classical techniques, the GPA advances quantum AI and sets the stage for future applications of quantum circuits in Reinforcement Learning (RL) robotics.

## I. Introduction

RL is a machine learning approach that allows autonomous intelligent agents to learn by directly interacting with an environment. These agents are rewarded for performing actions, and their goal is to find an optimal policy to maximize these rewards, which results in solving the task of the environment [1, 4, 38]. As artificial intelligence progresses, there is a greater need for algorithms that can learn rapidly and effectively, and speedups are more welcome than ever.

Quantum control refers to the use of classical or quantum RL agents to automatically design or optimize quantum circuits to address optimization tasks. The optimization objectives for the agent may include minimizing the number of gates, optimizing quantum states or entanglement, improving gate fidelity, and achieving other goals [1-14]. The agent uses optimization techniques to explore the vast design space of possible quantum circuits and selects designs that best meet the specified optimization objectives.

One set of metrologically useful states are squeezed states which give mild performance gains through pairwise entanglement generation [15]. In this case, a more complex sensor circuit is necessary to generate the quantum states that we need. The task therefore is to generate the specific kind of entanglement that will lead to an optimal quantum advantage for parameter estimation, for example, for the precision measurement of a phase shift. This phase shift could be generated by an inertial rotation, a magnetic field, or a variety of other possible terms in the system Hamiltonian of interest. The resulting technology will allow for a better understanding of the physical world with a breadth of applications that bridge many fields of science. We encapsulate this problem in a general conceptual framework referred to as a QSC [16, 17, 18].

A QSC executes a generalized Ramsey measurement on an array of qubits with encoding and decoding sequences represented by a chain of elementary gate operations, and the quantum design task is to select the optimal sequences [19]. The goal is to reveal the absolute quantum limit for measurement sensitivity when the circuit is taken as a whole. The design of such a QSC is difficult when the circuit is deep (meaning the cascade of many consecutive elementary gate operations) due to the large number of possible permutations of gates and measurements that should amplify the correct amplitudes for a sensitive signal while mitigating the adverse effects of noise and decoherence. While there are a variety of alternate approaches in optimal control theory, they all require a cycle of measurement and feedback, or open loop control, where exquisite modeling of the system is essential. This can lead to the degradation of the design utility when non-modelled perturbations are present [20]. These may include such imperfections as unitary errors due to gate inaccuracies, decoherence, dissipation, noise on control fields, and erasure errors of qubits.

In our proposed work, we developed a GPA for quantum control tasks. The GPA utilizes QPE and QPI to find an optimal policy that generates optimal QSCs with high QFI and less number of gates. The QFI is an essential quantum mechanical measure of precision and sensitivity within quantum parameter estimation [21, 22, 23, 24]. The GPA uses QPE for phase estimation to generate the search space and then applies QPI, which utilizes Grover search and amplitude amplification techniques to efficiently identify an optimal policy for generating optimal QSCs [7, 25, 26, 30]. Amplitude amplification is the process of increasing the probability of finding a desired solution by amplifying its amplitude through repeated steps of specific quantum operations, making the solution more likely to be measured [7, 30].

The contribution of the GPA is to generate optimal QSCs with high QFI values by generating entangled quantum states, specifically the squeezed states. This is achieved by employing the QPE and QPI to select the action, which corresponds to a quantum gate that maximizes the QFI of the generated QSCs while minimizing the number of gates. High QFI indicates that the circuit is more sensitive to parameter changes and therefore more informative or useful for quantum state estimation or other quantum control tasks. Few gates mean that the circuit is not complex and can be implemented in quantum computers.

The remainder of the paper is organized as follows. Section II explains the current state of the art in the field of Quantum Reinforcement Learning (QRL). Section III describes the methodology for the proposed GPA approach. Section IV describes the experimental results of the GPA for quantum control tasks. Finally, section V concludes the paper.

## II. RELATED WORK

Existing QRL approaches can be classified into two types: quantum agents that learn in classical environments and scenarios where the agent and environment are fully quantum.

### A. Quantum Agents in Classical Environments

Examples of the first type are found in [1, 2, 4, 6, 8, 10, 14]. Heimann et al. developed a Deep Quantum Reinforcement Learning (DQRL) for training a wheeled robot to navigate through complex environments [1]. The wheeled robot is a Double Deep Q-Network (DDQN) that interacts with an environment represented using Variational Quantum Circuits (VQCs). The authors used the data-reuploading method to transform the classical features into the VQCs. The robot was tested using three different scenarios of the Turtlebot 2 environment, such that the complexity of the environment increased in each scenario. The results of the proposed DQRL show that quantum machine learning can be applied for autonomous robotic enhancements.

Skolik et al. developed a Variational Quantum Algorithm based on Deep Q-Learning (VQA-DQL) for enhancing the training process of Parametrized Quantum Circuits (PQCs) that can be used to solve discrete and continuous state space RL tasks [2]. The authors tested the proposed QRL approach using the frozen lake and carte pole environments, and the results show that QRL can outperform classical RL in terms of generating q-values for better learning performance agents.

Samuel et al. developed a hybrid quantum-classical approach that consists of quantum circuits with tunable parameters to enhance the performance of Noisy Intermediate Scale Quantum (NISQ) devices [6]. The proposed approach consists of VQCs that represent classical DRL algorithms (e.g., experience replay and target network). The circuits represent a Quantum Neural Network (QNN) that is used to estimate the Q-value function, which is used to improve the decision-making and policy selection of NISQ systems by reducing energy costs.

Yun et al. presented a Centralized Training and Decentralized Execution Quantum Multi-Agent Reinforcement Learning framework (CTDE-QMARL), which addresses challenges related to NISQ and non-stationary properties [8]. The proposed framework incorporates innovative VQCs and demonstrates its effectiveness in a single-hop environment, improving the performance of rewards received by agents compared to classical frameworks.

Chen et al. introduced two deep quantum reinforcement learning frameworks: one utilizes amplitude encoding for the CartPole problem, while the other employs a hybrid Tensor-Network VQC (TN-VQC) architecture to handle high-dimensional inputs of the MiniGrid [10]. The results show the advantage of parameter saving with amplitude encoding and the potential for quantum reinforcement learning on NISQ devices through efficient input dimension compression.

Sequeira et al. presented a low depth policy for a reinforcement learning agent in a VQC [14]. The study demonstrates an efficient approximation of the policy gradient with logarithmic sample complexity relative to the number of parameters. Empirical results confirm the comparable performance of the proposed VQC policy gradient to classical neural networks in benchmarking environments and quantum control, utilizing few parameters, while also investigating the barren plateau phenomenon in quantum policy gradients through analysis of the fisher information matrix.

Dong et al. developed a QRL approach based on quantum superposition and parallelism for autonomous state value updating of agents [4]. The proposed QRL technique represents the action as a quantum superposition state, such that the eigenstate of the action is obtained by randomly observing the superposition state according to the collapse postulate of quantum measurement. The eigen action (eigen state) probability is determined by the probability amplitude and parallelly updated according to rewards. The proposed QRL provides a balance between exploration and exploitation and can speed up the learning process through quantum parallelism.

### B. Quantum Agents in Quantum Environments

Examples of quantum agents learning in quantum environments are found in [3, 5, 7, 9, 11, 12, 13]. Wei et al. presented a control RL algorithm called Deep Reinforcement Learning Quantum Experience Replay (DRL-QER) [7]. This quantum technique allows agents to receive experiences from the replay buffer according to the complexity and the replay times of the surrounding environment, which results in a balance between exploration and exploitation of their environment.

Alomari & Kumar presented a Reinforcement Learning Autonomous Quantum Agent (ReLAQA) that integrates a Grover Autonomous Quantum Agent (GAQA) with a Quantum TicTacToe (QTTT) game to outperform classical methods in action selection using quantum techniques [5]. The ReLAQA demonstrated faster and more efficient performance than classical techniques using fewer computational resources. This work paves the way for future developments in quantum circuits for reinforcement learning robotics and metrological enhancements in NISQ devices.

Wu et al. implemented a Quantum Deep Deterministic Policy Gradient (QDDPG) algorithm for efficient resolution of classical and quantum sequential decision problems [9]. The proposed QDDPG enables one-shot optimization for generating control sequences to achieve arbitrary target states, surpassing the need for optimization per target state as required by standard quantum control methods. Additionally, the algorithm facilitates the physical reconstruction of unidentified quantum states.

Wiedemann et al. proposed a QPE method that combines amplitude estimation and Grover search for solving quantum reinforcement learning tasks, achieving quadratically greater efficiency compared to classical Monte Carlo methods [11]. Using QPE, the authors developed a QPI approach that iteratively improves policies using the Grover search. The authors provide implementation and simulation for a two-armed bandit markov decision process to showcase the effectiveness of the proposed approach.

Borah et al. developed a Deep Reinforcement Learning Artificial Neural Agent (DRLANA) to control highly nonlinear quantum systems toward the ground state [12]. By incorporating weak continuous measurements into the proposed DRLANA, successfully learns counterintuitive strategies and achieves high fidelity. This approach demonstrates effective control techniques for nonlinear Hamiltonians, surpassing traditional methods.

Sivak et al. proposed a Model-Free Circuit-based Reinforcement Learning (MFCRL) approach for training an agent on quantum control tasks, addressing the issue of model bias [13]. The agent learns the parameters of a control PQC through trial-and-error interaction with the quantum system, utilizing measurement outcomes as the sole source of information. The proposed approach enables rewarding the agent using experimentally available observables, facilitating the preparation of nonclassical states, and executing logical gates on encoded qubits.

Dong et al. proposed the Quantum-inspired Reinforcement Learning (QiRL) algorithm for autonomous mobile robot navigation control [3]. The proposed technique uses a probabilistic action selection technique and a reinforcement policy, which are inspired, respectively, by the quantum measurement collapse phenomenon and amplitude amplification.

TABLE I
QRL APPROACHES

| Author | Approach | Environment Type | Advantages | Disadvantages | Metrics |
|---|---|---|---|---|---|
| Heimann et al. [1] | DQRL | Classical environment | Enhances the learning process of autonomous robotics applications | Qubit decoherence. The trainable parameters decrease the learning performance | Training steps, the number of trainable parameters, and average reward |
| Skolik et al. [2] | VQ ADQL | Classical environment | Provide an efficient decision policy | Qubit decoherence. Not a fully quantum agent. High complexity | Error median and Training steps |
| Samuel et al. [6] | RLVQCs | Classical environment | Enhances the learning process of VQCs | The implementation of the agent is not fully quantum. Classical computational errors in estimating the hyperparameters | Fidelity and accuracy |
| Yun et al. [8] | CTDE-QMARL | Classical environment | Limited applicability to complex domains | Improves the performance of rewards | Average reward |
| Chen et al. [10] | TN-VQC | Classical environment | Efficient data dimension compression | High complexity | Average reward |
| Wiedemann et al. [11] | QPE | Quantum environment | Efficient learning | Limited scalability for complex environments | Average reward |
| Sequeira et al. [14] | VQC | Classical environment | Efficient learning | Does not provide a balance between exploration and exploitation | Average reward |
| Dong et al. [4] | QRL | Classical environment | Provides a balance between exploration and exploitation | Not a fully quantum agent. Deals with small state space, which makes the learning process insufficient | Training steps |

| Author | Approach | Environment Type | Advantages | Disadvantages | Metrics |
|---|---|---|---|---|---|
| Alomari & Kumar [5] | ReLAQA | Quantum environment | Provides a balance between exploration and exploitation | There is no phase estimation to generate efficient search space | Observed states |
| Wei et al. [7] | DRL-QER | Quantum environment | Provides a balance between exploration and exploitation | Qubit decoherence. Not a fully QRL approach | Accuracy and precision |
| Wu et al. [9] | QDDPG | Quantum environment | Efficient one-shot optimization | Limited scalability for large-scale problems | Average reward |
| Borah et al. [12] | DRLANA | Quantum environment | Enhances quantum state control | High computational cost | Fidelity |
| Sivak et al. [13] | MFCRL | Quantum environment | Eliminates model bias | High complexity | Fidelity |
| Dong et al. [3] | QiRL | Quantum environment | Efficient learning performance | The representation of the state space is not accurate | Accuracy and learning rate |

Table I shows the surveyed QRL approaches and their limitations. In the proposed work, we developed a GPA for generating optimal QSCs capable of overcoming the limitations discussed in Table I and solving complex quantum physics problems. Our proposed GPA differs from the methods in [1-14] as it is not an enhanced quantum version of classical techniques; instead, it is a pure quantum agent that uses phase estimation, Grover search, and amplitude amplification techniques to maximize the QFI of the generated QSCs while minimizing the number of gates.

TABLE II
THE GPA VS RELATED QUANTUM AGENTS FOR QSCS

| Feature | GPA | QRL [4] | QPE [11] | ReLAQA [5] |
|---|---|---|---|---|
| Optimization Goal | Maximizes QFI while minimizing the number of gates for QSCs | Balances exploration and exploitation, speeding up learning via quantum parallelism | Increases efficiency in policy evaluation through Grover search and amplitude amplification | Focuses on action selection efficiency, using fewer resources than classical techniques |
| Efficiency in Gate Selection | Achieves high QFI with fewer gates, by using the QPE to minimize the search space | Uses quantum superposition for state updating but does not focus on minimizing the number of gates for QSCs | Implements QPI with Grover search but does not aim to minimize gate count specifically in QSCs | Utilizes efficient action selection technique but not optimized for QFI or minimal gate usage in QSCs |
| Target Outcome | Generates entangled squeezed states, which enhances sensitivity for QSCs | Produces action as quantum superposition for faster learning, not focused on producing entangled states for QSCs | Primarily addresses QPI efficiency, not focused on producing entangled states for QSCs | Enhances action selection in quantum games, not focused on producing entangled states for QSCs |
| Scalability and Practicality | Shows scalability in QSCs for complex quantum physics problems | Enhances learning in quantum agents but less scalable for QSCs | Quadratic efficiency gain compared to classical approaches but less scalable for QSCs | Applicable to quantum games and NISQ devices but less scalable for QSCs |

Table II compares the GPA with other QRL approaches, illustrating why GPA is more effective for designing optimal QSCs. Unlike other methods in [4, 5, 11], which focus on state updating, policy improvement, or action selection in quantum environments, GPA is specifically tailored to optimize QSCs by maximizing QFI while minimizing the number of gates. This approach ensures higher sensitivity for quantum state estimation and more efficient resource use in circuit design.

III. METHODOLOGY

Our proposed QRL approach consists of the GPA and the QSC environment represented. Fig. 1 illustrates the workflow of the proposed GPA approach.

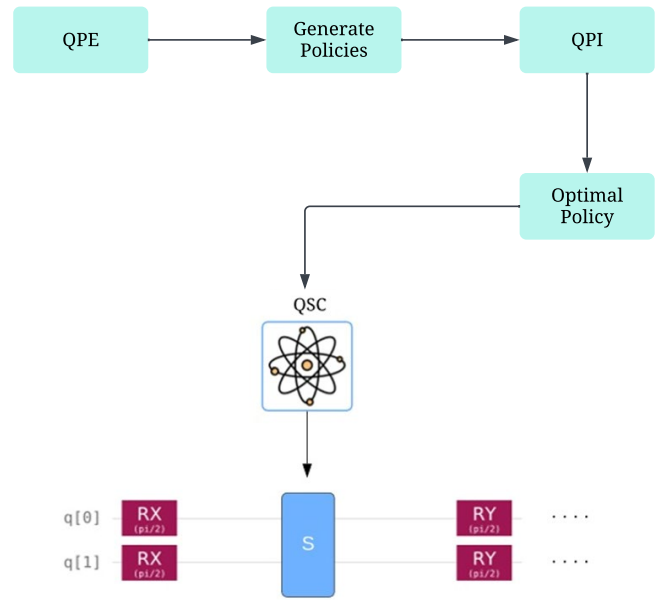

Figure 1. The Proposed GPA Workflow

As shown in Fig. 1, The workflow of the proposed GPA consists of the QPE and the QPI. The idea of the QPE is to generate the search space, which consists of the evaluated policies. The idea of the QPI is to apply Grover search and amplitude amplification techniques to the search space to find an optimal policy that will be used to generate QSCs.

*A. The QSC Environment*

To illustrate the efficient performance of the proposed GPA, we consider a QSC that consists of two qubits and $R_x$, $R_y$, and $S$ gates. Fig. 2 shows the structure of the QSC. This QSC has an optimal solution that we intend the GPA to generate, namely it should produce the N00N state since that maximizes the QFI [27].

First, we initialize the qubits to the state $|0\rangle$. Then, we apply a generalized Ramsey sequence to measure the QFI, which represents the accumulation of a relative phase between the two qubits collective dipole and a stable local oscillator that synchronizes the timing. This phase is associated with the generator of the Ramsey sequence such as the Pauli-Z operator ($Z$) [28, 29]. The accumulation consists of a sequence of $R_x$, $R_y$, and $S$ gates that the GPA optimally generates to maximize the QFI. The QFI and $S$ are given in equations 1 and 2 as follows.

$$QFI = \frac{4(\langle\psi|Z^2|\psi\rangle - \langle\psi|Z|\psi\rangle^2)}{n} \quad (1)$$

$n$ is the number of the qubits in the QSC, which is 2. $|\psi\rangle$ is the quantum state that is manipulated using the quantum gates that are generated by the GPA. $Z$ is the generator related to the rotation angle $\theta$. $Z$ represents the projection of the angular momentum of a quantum state along the z-axis by $\theta = \frac{\pi}{2}$. It is used to measure the z-component of the quantum state angular momentum.

$$S = \exp(-i\theta Z^2) \quad (2)$$

To calculate the QFI using a fully quantum approach, without relying on classical techniques, we extended the QSC to incorporate 4 qubits. We then developed a quantum subtraction operation to represent the QFI. This operation computes the difference between the sum of qubits 0 and 1 and the sum of qubits 2 and 3. An additional 2 qubits were used to execute this subtraction process, resulting in a 6-qubit circuit, as illustrated in Fig. 2.

Fig. 2 represents the extended version of the 2-qubit QSC that incorporates the QFI subtraction operation. The circuit consists of 6 qubits: qubits 0-3 represent the QSC with all the left part gates, which are $R_x$, $R_y$, $R_z$, and squeezing ($S$) gates. The rotation angles of $R_x$ and $R_y$ are $\frac{\pi}{2}$ and the rotation angle of $R_z$ is 0.05. Qubits 4-5 are counting qubits to account for the borrow bits, which represent the sign of the subtraction results. The inverse of the CDKM Ripple Carry Adder ($adder\_inv$), which is a package used in qiskit is applied to perform the subtraction operation between the subtrahend (qubits 0 and 1) and the minuend (qubits 2 and 3). The $adder\_inv$ adder utilizes controlled-not ($CNOT$) and Toffoli gates to handle carry and borrow operations, ensuring any borrow needed is properly accounted for as the subtraction proceeds. See Table III in the appendix for more information about these gates. Borrow information and intermediate results are stored in qubits 4-5, which represent the sign of the final result. The subtraction operation is completed by entangling these intermediate results and performing a final controlled operation to yield the result. The outcome represents the QFI, which is the subtraction of the subtrahend from and the minuend, with qubits 4-5 indicating the sign of the result. By integrating the proposed subtraction operation, we ensure that the QFI is computed through a quantum approach, where its generator is a rotation around the z-axis using the $R_z$ gate. The results of this QSC are shown in Figs. 4-6.

*B. The QPE*

The idea of the QPE is to generate the search space using phase estimation, while the QPI utilizes Grover search and amplitude amplification techniques over the search space to identify an optimal policy that generates optimal QSCs. Fig. 3 represents the phase estimation circuit of the QPE. It consists of four counting qubits $|0\rangle$, four Hadamard ($H$) gates, four controlled-unitary ($CU$) gates with rotation angles of $\frac{\pi}{4}$, Quantum Fourier Transform (QFT), four measurement gates, and the environment $|\psi^\phi\rangle$ that represents the QSC in Fig. 2 [30, 31]. The output of the circuit X represents the value function of each estimated policy of the environment.

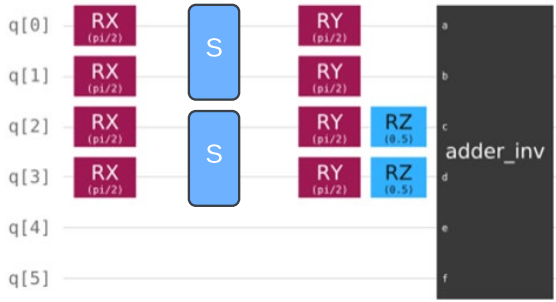

Figure 2. The Proposed QSC

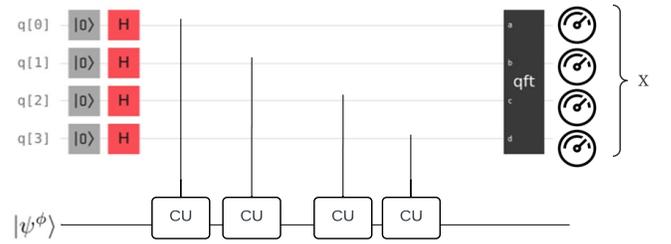

Figure 3. The Proposed QPE Circuit

The QPE circuit performs a phase estimation on $|\psi^\phi\rangle$ to generate the search space, which is the evaluated policies. These policies are generated by evaluating their value functions using the following equation.

$$v_\pi^H(s) = \mathbb{E}_{t^H}[G(t^H)|S_0 = s] \quad (3)$$

Where $s$ is the initial state, the return $\mathbb{E}$ is taken concerning the distribution of all trajectories of length $H$. $G$ is the generator of the QSC that represents a Pauli-Z operator.

Evaluating $v_\pi^H(s)$ the complexity of directly calculating the expectation grows exponentially in the horizon H.

The first step of the QPE is to encode $v_\pi^H(s)$ as an amplitude of a basis vector in a quantum superposition. Assume that we know a lower bound $\underline{g}$ and an upper bound $\overline{g} \neq \underline{g}$ on all returns. Using the function $\phi(x) = (x - \underline{g})/(\overline{g} - \underline{g})$, we can say that $v_\pi^H(s)$ can be presented by the following unitary operator $\Phi$, where $|x\rangle$ is a policy.

$$|x\rangle|0\rangle \xrightarrow{\Phi} |x\rangle\left(\sqrt{1-\phi(x)}|0\rangle + \sqrt{\phi(x)}|1\rangle\right) \quad (4)$$

The QPE can efficiently estimate the value function $v_\pi^H(s)$ of each policy using the following equation, where $A_{QPE}^\phi = (id_{T^H} \otimes \Phi) \circ A_{QPE}^\phi$.

$$Q_{QPE} = -A_{QPE}^\phi \circ S_0 \circ \left(A_{QPE}^\phi\right)^\dagger \circ (id_{T^H \otimes g} \circ Z) \quad (5)$$

$\Phi$ acts on the return subsystem and the counting qubits. $S_0$ is a phase operator that flips the phase of a state when the trajectory, the return, and the counting qubit are in $|0\rangle$. $Z$ is the Z gate that flips the phase of a state around the z-axis when the counting qubit in $|1\rangle$.

*C. The QPI*

The QPI consists of the Grover search and amplitude amplification. It will perform a Grover search on the output generated by the QPE to find an optimal policy that will be used to generate quantum QSCs. For a fixed policy $\pi$, QPE can be represented as one unitary operation that maps $|x\rangle|0\rangle \xrightarrow{QPE} |\psi_{QPE}^\pi\rangle$. To implement the QPI, we need to prepare the search space that is generated by the QPE using the following equation.

$$A_{QPI} = \left(\prod_{\pi \in P} OPE_\pi\right) \circ (H_P \otimes id) \quad (6)$$

$OPE_\pi$ represents the value function that is generated by the QPE for a policy $\pi \in P$, where $P$ represents all the generated policies. After generating the search space using $A_{QPI}$, we implement an oracle and use it to amplify the amplitudes of policies that have higher returns. For a policy $\pi$ and its estimated value function $v_\pi$, the oracle $O_\pi$ is given by the following equation, where $t$ is the number of the qubits in the QPE circuit.

$$|x\rangle \xrightarrow{O_\pi} \begin{cases} -|x\rangle & \text{if } \frac{1}{\phi}(sin^2(\pi x/2^t)) > v_\pi \\ |x\rangle & \text{else} \end{cases} \quad (7)$$

Using the oracle $O_\pi$ that will target the optimal policy, we represent the QPI by the following equation.

$$O_{QPI}^v = -A_{QPI} \circ S_0 \circ A_{QPI}^\dagger \circ (id_P \otimes O_\pi) \quad (8)$$

Based on the concept of amplitude amplification that is used in the Grover search, applying $O_{QPI}^v$ a certain number of times to the search space generated by $A_{QPI}$ will amplify the amplitude of the states that satisfy the condition $\frac{1}{\phi}(sin^2(\pi x/2^t)) > v_\pi$ [11, 26, 32, 33]. This will result in increasing the probability of measuring a policy $\pi$ that has better performance than $v_\pi$.

To ensure the effectiveness of the amplitude amplification, it is essential to define the number of Grover rotations. This indicates how many times we run the QPE, which determines the scaling of amplitudes for the desired states. While there exists an efficient number of Grover rotations in theory, determining this value in QPI is difficult due to the necessity of knowing the probability of achieving the desired state upon measuring the current state [19, 34, 35]. To select the number of Grover rotations $L$, we used $L = int(k(r + V(s')))$. This method takes in the reward $r$ received by the GPA and the estimated value function $V(s')$ of the new state $s'$ visited by the GPA and returns $L$ as an integer number. $k$ is a learning hyperparameter that shows that $L$ is proportional to $(r + V(s'))$ [4, 5]. Here, $sin(2L + 1)\theta$ is a periodical function around $(2L + 1)\theta$. Too many iterations may generate small probability amplitudes, then we select $L = min[int(k(r + V(s'))), int(\frac{\pi}{4\theta} - \frac{1}{2})]$. This ensures that $L$ remains within an optimal range, which efficiently maximizes the probability amplitude. See section IV for more detailed information.

IV. EXPERIMENTAL RESULTS

We implemented the proposed GPA using local simulation. The simulation is done on an Intel Core I-7 Dell computer with 16 GB of RAM using Qiskit through Python. All experiments use the statevector simulator. Furthermore, we compared the performance of the GPA with the GAQA, which is a quantum agent that we developed, utilizing Grover search and amplitude amplification [5]. The main difference between the proposed GPA and the GAQA is that the GPA uses phase estimation to generate the policies, which reduces the search space. Also, the GPA generates the QFI using the subtraction operations in Figs. 2 and 7. These enhancements are not implemented in the GAQA. The results illustrate the efficient performance of the proposed GPA in generating optimal QSCs compared to the other approaches.

*A. The Performance of the Proposed GPA*

To demonstrate the computational power of the proposed GPA we integrated the QSC in Fig. 2 into the QPE phase estimation circuit as $|\psi^\phi\rangle$ to generate the policies, and then the QPI was applied to find the best policy that represents the optimal QSC in Fig. 1. Figs. 4, 5, 6, and 8 show the experimental results that are generated using the statevector simulator.

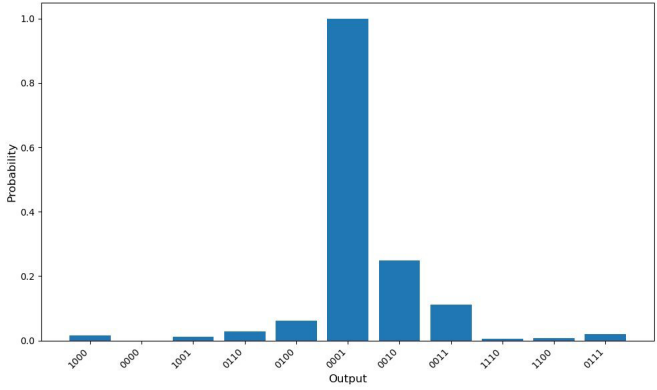

Figure 4. The Results of the QPE

Fig. 4 shows the distribution of the QPE outputs for the value function $v_\pi^H(s)$. We executed the QPE phase estimation circuit that is shown in Fig. 3 for 4096 shots, resulting in an evaluated value of $v_\pi^H(s) = 0.99$. The circuit consists of 12 qubits, with 4 counting qubits and the remaining 8 qubits represent the QSC $|\psi^\phi\rangle$. The four counting qubits represent the four bits, where each four bits represents a phase. We measure only the computing qubits to estimate the value function of each policy. Achieving a high $v_\pi^H(s)$ of 0.99 means that the QPE provides a quantum advantage in estimating the value functions of the policies, making it suitable for estimating the phase of complex search spaces.

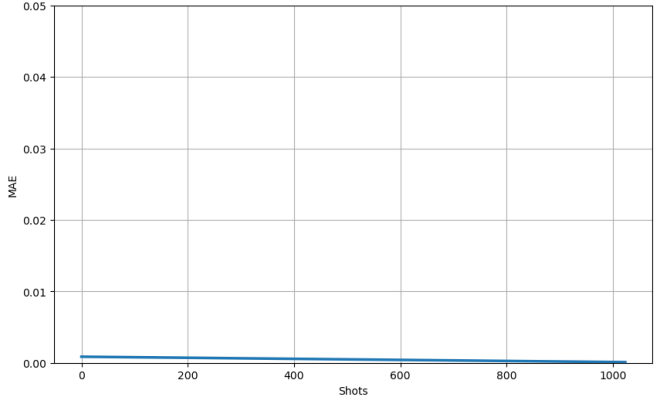

Figure 5. The MAE of the QPE

Fig. 5 illustrates the Median Approximation Error (MAE) of the QPE over 1024 shots, where each shot represents an execution of the QPE circuit. The error consistently decreases as the number of shots increases, showcasing the quantum advantage of the QPE algorithm in accurately estimating the phase. This illustrates the robustness and efficiency of the QPE in generating complex search spaces.

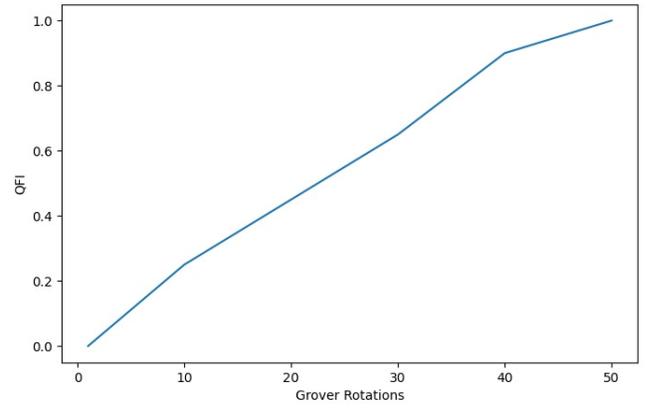

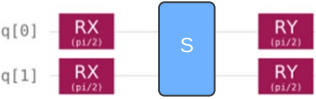

Figure 6. The Results of the QPI

Fig. 6 describes the results of running the QPI over the policies that were generated by the QPE. The QPI was able to find the optimal policy that represents the QSC in Fig. 1. The generated QSC represents an optimal design for generating the N00N state because less complex in terms of the number of gates and maximizes the QFI by generating a QFI of 1 [27]. We used a range between 0 and 1 to represent the QFI values, bringing them to a standard format to enhance clarity for readers. The QPI used 50 Grover rotations, where each rotation represents an amplitude amplification, as described in equation 8. We did not use any methods to select the number of Grover rotations, instead, we conducted several experiments with different numbers of Grover rotations and found that 50 rotations generate a high QFI of 1.

The previous experiment of the GPA is considered as one episode. To run the GPA for multiple episodes we simplified the QSC by reducing the number of the qubits, eliminating the squeezing gate, and simplifying the subtraction operation that generates the QFI. Fig. 7 shows the simplified QSC, which allows us to handle the high complexity of the search space and makes it possible to run the GPA for two episodes [4].

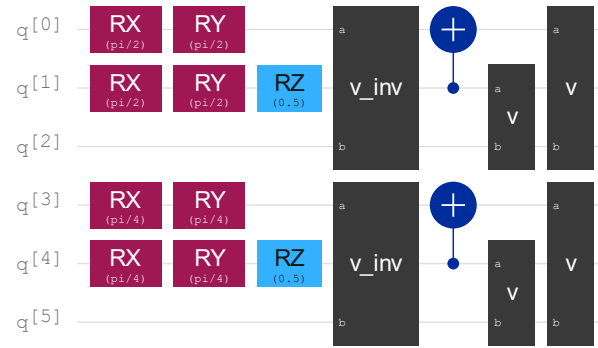

Figure 7. The Simplified QSC

Fig. 7 shows the simplified QSC. It consists of two QSCs, where each represents an episode. The first QSC 1 consists of qubits 0-2, representing the first episode, with $R_x$ and $R_y$ gates having rotation angles of 90 degrees. The second QSC 2 consists of qubits 3-5, representing the second episode, with $R_x$ and $R_y$ gates having rotation angles of 45 degrees. The rotation angle of the $R_z$ gate has a range from 0 to 0.1, with 0.05 used as an example to illustrate the circuit. The rotation angle increases through the Grover rotations as we aim to measure the QFI of both circuits concerning the generator $R_z$. Each circuit has one $CNOT$ gate, two controlled-v ($CV$) gates, and one inverse controlled-v ($CV_{inv}$), which they and the $R_z$ gate represent the simplified subtraction operation for generating the QFI of each circuit. The $V$ and the inverse $V_{inv}$ gates were developed in [36]. The $V$ gate applies the operation $V = \frac{1+i}{2}\begin{pmatrix} 1 & -i \\ -i & 1 \end{pmatrix}$ to a qubit, while the $V\_inv$ applies the operation $V_{inv} = \frac{1-i}{2}\begin{pmatrix} 1 & i \\ i & 1 \end{pmatrix}$ to a qubit. For more details about the subtraction operation that is used in Fig. 7 and the $CV$ and $CV_{inv}$ see [36, 37]. We are using controlled versions $CV$ and $CV_{inv}$, which they maintain the properties of the $V$ and $V\_inv$ gates, but they activated and deactivated using a control qubit. The goal of the $CV$ and $CV_{inv}$ gates to hold the borrow bit, which acts as the sign bit of the subtraction operation [37]. The goal of the GPA is to find the optimal QSC that has the highest QFI. The results are shown in Fig. 8.

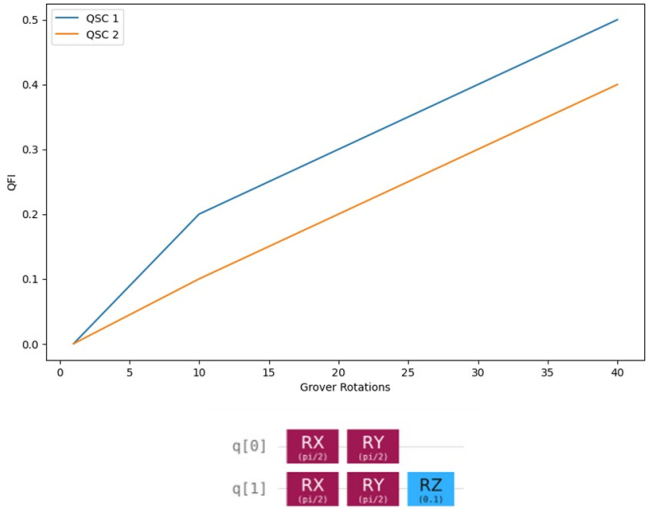

Figure 8. The Results of the Simplified QSC

Fig. 8 illustrates the results of running the GPA on the simplified QSC that consists of QSC 1 and QSC 2. First, the QPE generated the search space, and then we applied the QPI to find the optimal QSC which represents QSC 1. Since there are no squeezing $S$ gates in both QSC1 and QSC 2, the maximum QFI of the circuits will be 0.5. This is because the $S$ gate generates entangled squeezed states that maximize the QFI, which increases the sensitivity of the QSC. In the beginning, both QSCs generated a QFI of 0 because the rotation angle of $R_z$ was 0. As Grover rotations increased, the QFI value for both circuits increased, but as seen in Fig. 8 QSC 1 has a higher QFI. At Grover rotation 40, QSC 1 generated a QFI of 0.5, and the GPA selected this circuit as the optimal QSC, printing it with an $R_z$ rotation angle of 0.1. QSC 2 was not selected as the optimal QSC, because it has less QFI values during all of the Grover rotations. This illustrates the quantum advantage of the proposed GPA in terms of generating optimal QSCs.

### B. The Results of Comparing the GPA with the GAQA

We compared the performance of the GPA with the GAQA, which is a quantum agent that we developed, utilizing Grover search and amplitude amplification [5]. The main difference between the proposed GPA and the GAQA is that the GPA uses phase estimation to generate the policies by estimating their value function. This reduces the search space and efficiently allows the QPI to find the best policy that represents an optimal QSC. Additionally, the GPA executes the two episodes shown in Fig. 7 in parallel, unlike the GAQA, which runs each episode sequentially. Fig. 9 shows the workflow of the GAQA.

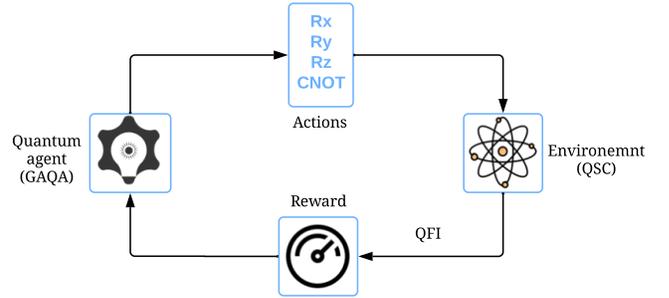

Figure 9. The Workflow of the GAQA

Fig. 9 shows the workflow of the GAQA. In our previous work, we used the GAQA to solve complex robotics applications, such as the QTTT environment [5]. The problem is to find the optimal quantum circuit that solves the QTTT. In this experiment, we modified the actions to those shown in Fig. 9. We want the GAQA to find the optimal QSC that is represented by qubits 0-2 in Fig. 8. The GAQA starts by selecting an action, which is a gate that can be $R_x$, $R_y$, $R_z$, and $CNOT$. It then generates the QFI of the selected action using equation 1. Next, the GAQA utilizes amplitude amplification to increase the probability amplitude of the selected action. It then uses this probability and the generated QFI as a reward to select the next action that maximizes the QFI. This process continues until the GAQA generates a QFI of 1 or reaches the number of actions allowed in an episode which is 10. Modifying the GAQA is not a straightforward application since we had to change the actions to generate optimal QSCs. For more detailed information about the GAQA see [5]. To ensure a fair comparison, we ran the GAQA for two episodes, matching the GPA limitation of only two episodes, which is important for managing the high complexity of the search space. The results are shown in Fig. 10.

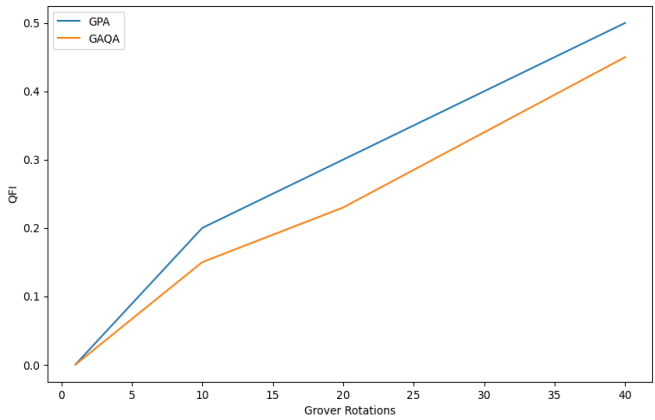

Figure 10. The QFI of the GPA vs GAQA

Fig. 10 illustrates the QFI of the GPA compared to the GAQA over the simplified QSC in Fig. 7. The GPA generated a QFI of 0.5, outperforming the GAQA, which generated a QFI of 0.45. Both agents generated a QFI of 0.0 at Grover rotation 1 because the rotation angle of $R_z$ was 0. As Grover rotations increased GPA started to generate higher QFI values until it reached a QFI of 0.5 at Grover rotation 40. Moreover, the GAQA took 40 Grover rotations and did not reach a QFI of 0.5. This demonstrate the quantum advantage of the proposed GPA in terms of generating optimal QSCs.

## V. Conclusion

Quantum control involves the utilization of classical or quantum RL agents for designing and enhancing quantum circuits to address optimization challenges. The optimization objectives for the agent may include minimizing the number of gates, optimizing quantum states or entanglement, improving gate fidelity, and achieving other goals. In this proposed work, we have developed a GPA approach for generating optimal QSCs capable of solving complex quantum physics problems. The QSC performs a generalized Ramsey measurement on qubits using sequences of quantum gates, and the task of the GPA is to select the optimal sequences. The proposed GPA consists of the QPE and the QPI. The idea of the QPE is to generate the search space, which consists of the evaluated policies. QPI then uses Grover search and amplitude amplification techniques to find an optimal policy that will be used to generate QSCs with high QFI and few quantum gates. High QFI indicates that the circuit is more sensitive to parameter changes and therefore more informative or useful for quantum state estimation or other quantum control tasks. Few quantum gates mean that the circuit is not complex and can be implemented in quantum computers.

To evaluate the performance of the proposed GPA, we considered a QSC that consists of two qubits and a sequence of $R_x$, $R_y$, and $S$ gates. This circuit has an optimal N00N state that maximizes the QFI, and the task of the GPA is to generate this circuit while minimizing the number of gates. The results show the efficient performance of the proposed GPA by generating optimal QSCs with a QFI of 1. Next, we simplified the QSC to enable the GPA to run for multiple episodes, and then we compared its performance with the GAQA. The results show that the GPA outperforms the GAQA by generating higher QFI values. The implementation details and simulations we conducted will illustrate how quantum agents can be utilized to solve quantum physics problems.

The practical application of the GPA extends to quantum sensing and metrology, fields where critical measurements are essential. High sensitivity and fidelity are necessary to measure physical parameters accurately, and by generating circuits with high QFI and few gates, the GPA supports the development of efficient and scalable QSCs. This can significantly enhance sensitivity in measurement systems and advance metrological precision, pushing the boundaries of what can be detected or quantified in quantum systems. For future work, we intend to evaluate the performance of the proposed GPA on different quantum physics problems, which require the design of complex QSCs. Moreover, we intend to explore Hybrid Classical-Quantum agents (HCQAs) methods that combine classical algorithms with quantum techniques to enhance decision-making processes and optimize the design of QSCs.

## Acknowledgments

This work was supported by the National Science Foundation Grant No. OMA 2231377.

APPENDIX

TABLE III
NOTATION DEFINITION

| Notation | Definition |
|---|---|
| $QFI$ | Represent the sensitivity of the QSC |
| $\lvert\psi\rangle$ | Quantum state in the $QFI$ |
| $n$ | The number of qubits in the $QFI$ |
| $v_\pi^H(s)$ | The value function of each policy |
| $H$ | The length of the trajectories in $v_\pi^H(s)$ |
| $Id$ | The identity matrix |
| $Q_{QPE}$ | The QPE that uses phase estimation to generate the search space |
| $\Phi$ | The unitary operator that represents $v_\pi^H(s)$ in the QPE |
| $O_{QPI}^v$ | The QPI that uses Grover search and amplitude amplification to find the optimal QSC |
| $O_\pi$ | The oracle in the QPI |
| 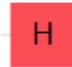 | Hadamard gate acts on one qubit |
| 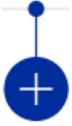 | Controlled-not gate acts on two qubits |
| 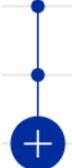 | Toffoli gate acts on three qubits |
| 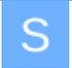 | Squeezing gate acts on two qubits, which creates entangled squeezed states |
| 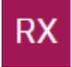 | The $R_x$ gate represents a rotation around the x-axis by a rotation angle |
| 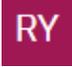 | The $R_y$ gate represents a rotation around the y-axis by a rotation angle |
| 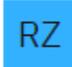 | The $R_z$ gate represents a rotation around the z-axis by a rotation angle |
| 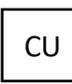 | Controlled-unitary gate acts on multiple qubits |
| 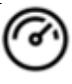 | Measurement gate acts on one qubit |
| 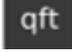 | The QFT that is used in the QPE |